\documentclass{aa}  
\usepackage{comment}
\usepackage{url}
\usepackage[colorlinks,citecolor=blue]{hyperref}\usepackage{graphicx}
\usepackage{natbib}
\usepackage{natbib}
\bibpunct{(}{)}{;}{a}{}{,} % to follow the A&A style
\usepackage{txfonts}
\begin{document}

   \title{Angular momentum transport via gravitational instability in the Elias 2-27 disc}

   \subtitle{}

   \author{C. Longarini\inst{1,2}, G. Lodato\inst{2}, C. J. Clarke\inst{1}, J. Speedie\inst{3}, T. Paneque-Carreño\inst{4,7}, E. Arrigoni\inst{2}, P. Curone\inst{2}, C. Toci\inst{4}, C. Hall\inst{5,6}
          } 

   \institute{Institute of Astronomy, University of Cambridge, Madingley Road, Cambridge, CB3 0HA, United Kingdom\\
    \email{cl2000@cam.ac.uk}
   \and 
   {Dipartimento di Fisica, Università degli Studi di Milano, Via Celoria 16, Milano, 20133, Italy}
   \and
   {Department of Physics \& Astronomy, University of Victoria, Victoria, BC V8P 5C2, Canada}
   \and
   {European Southern Observatory, Karl-Schwarzschild-Strasse 2, D-85748 Garching bei Munchen, Germany}
   \and
   {Department of Physics and Astronomy, The University of Georgia, Athens, GA 30602, USA}
   \and 
   {Center for Simulational Physics, The University of Georgia, Athens, GA 30602, USA}
   \and 
   {Leiden Observatory, Leiden University, P.O. Box 9513, NL-2300 RA Leiden, the Netherlands}
   }

   \date{Received ; accepted }
\titlerunning{Angular momentum transport in  Elias 2-27}
\authorrunning{Longarini et al.}
% \abstract{}{}{}{}{} 
% 5 {} token are mandatory

  \abstract
   {Gravitational instability is thought to be one of the main drivers of angular momentum transport in young protoplanetary discs. The disc around Elias 2-27 offers a unique example of gravitational instability at work. It is young and massive, displaying two prominent spiral arms in dust continuum emission and global non-axisymmetric kinematic signatures in molecular line data.

   In this work, we used archival ALMA observations of $^{13}$CO line emission to measure the efficiency of angular momentum transport in the Elias 2-27 system through the kinematic signatures generated by gravitational instability, known as 'GI wiggles'. Assuming the angular momentum is transported by the observed spiral structure and leveraging previously-derived dynamical disc mass measurements, the amount of angular momentum transport we found corresponds to an $\alpha-$viscosity of $\alpha=0.038\pm0.018$. This value implies an accretion rate onto the central star of $\log_{10}\dot{M}_\star=-6.99\pm0.17\text{M}_\odot/\text{yr, which}$  reproduces the one observed value of $\log_{10}\dot{M}_{\star,\text{obs}}=-7.2\pm0.5\text{M}_\odot/\text{yr }$ very well. The excellent agreement we have found serves  as further proof that gravitational instability is the main driver of angular momentum transport acting in this system.%  {The relatively high value of the $\alpha-$viscosity we found implies that in young discs the transport of angular momentum is more efficient}
}


   \maketitle
%
%-------------------------------------------------------------------

\section{Introduction}

Elias 2-27 is a young ($\sim0.5$Myr, $M_\star =0.46\text{M}_\odot$) M0 star \citep{andrews09} located at a distance of 116 pc \citep{Eliasdistance} in the $\rho-$Oph star forming region, hosting a likely gravitationally unstable disc. The circumstellar disc shows two large-scale trailing spiral arms in dust continuum emission \citep{Perez16}, whose origins were first attributed to gravitational instability, due to the high dust mass. An estimate of the total dynamical mass of the disc has been provided by \cite{Veronesi21}, $M_d = 0.08\pm0.04\text{M}_\odot$, implying a disc-to-star mass ratio of 17\%. \cite{Meru17} performed three-dimensional (3D) numerical SPH simulations to investigate the origin of the spiral structure: by comparing gravitational instability and internal and external companion scenarios, they found that GI best reproduces the observed morphology. Similar results have also been found by \cite{Hall18}. In addition, due to its high brightness, Elias 2-27 has become part of the the DSHARP sample \citep{Dsharp1}, allowing for more thorough studies of its dust morphology. \cite{Dsharp3} characterised annular substructures in the discs within the DSHARP sample, finding that Elias 2-27 has a gap at $\sim 70\text{au}$. Even though the main focus of the DSHARP program was dust emission, also kinematic data on CO isotopologues were collected. \cite{Pinte20} found complex kinematic features in the Elias 2-27 system, showing perturbations to the velocity field. However, due to the low resolution of the data, a detailed analysis was not possible. \cite{paneque21} presented new data on this system and conducted a detailed analysis of the morphology and the kinematics. Global perturbations in the velocity field of $^{13}$CO and C$^{18}$O  were found and their morphology follows the shape of the spiral, consistent with kinematic signatures of GI induced density waves, or GI Wiggles \citep{Hall20,Longarini21, Terry22}. From the same dataset, \cite{Veronesi21} studied the rotation curve and obtained the previously mentioned disc mass, by measuring the super-Keplerian contribution of the disc self-gravity.

In this letter, we study the GI Wiggle in Elias 2-27. Thanks to the disc mass estimate provided by \cite{Veronesi21}, we have been able to constrain the amount of angular momentum transported throughout the disc. In Section 2, we discuss the theoretical framework we use in this Letter. In Section 3, we present the dataset and the analysis. In Section 4, we discuss the results and in Section 5 we draw our conclusions.

\section{Gravitational instability in protoplanetary discs}
%\subsection{Gravito-turbulence and GI wiggles}
The onset of gravitational instability is determined by the Toomre parameter
\begin{equation}
    Q=\frac{c_s\kappa}{\pi G \Sigma}, 
\end{equation}
where $\kappa$ is the epicyclic frequency, which (for a Keplerian disc) is just the Keplerian frequency, $\Omega_k$; $\Sigma$ is the surface density; and $c_s$ the sound speed. Essentially, $Q$ measures the strength of stabilising terms (e.g. pressure and rotation, at the numerator) compared to destabilising ones (e.g. self-gravity, at the denominator). A protoplanetary disc is marginally unstable when $Q\simeq 1$: in this regime, the disc develops a spiral structure and, by means of shocks, this leads to energy dissipation and heating. One of the most important consequences of gravitational instability is its ability to drive angular momentum transport throughout the disc, and therefore induce accretion onto the central object \citep{LBK72}. 

\subsection{Gravito-turbulence and transport of angular momentum}

A disc is in the gravito-turbulent regime when its angular momentum transport is driven by gravitational instability. In this context, the $R\phi$ component of the vertically integrated stress tensor can be written as:
\begin{equation}
    T_{R\phi} = \int \left\langle\frac{g_Rg_\phi}{4\pi G} \right\rangle \text{d}z,
\end{equation}
where $g_R,g_\phi$ are the radial and azimuthal component of the perturbed self-gravitating field, and the brackets indicate azimuthal averaging. To this stress, it is necessary to  also add the induced Reynolds
stress, given by:
\begin{equation}
    T_{R\phi} = \Sigma \langle \delta v_R v_\phi \rangle,
\end{equation}
where $\delta v_R$ and  $\delta v_\phi$ are the perturbed fluid velocities \citep{balbus99}.  In the classic viscous scenario, this term is responsible for angular momentum transport in the accretion disc. The Shakura and Sunyaev $\alpha-$prescription \citep{SS73} relates the stress tensor to the local disc pressure 
\begin{equation}
    T_{R\phi} = \alpha \Sigma c_s^2 \frac{\text{d}\log \Omega}{\text{d}\log R},
\end{equation}
where, for a Keplerian disc, ${\text{d}\log \Omega_k}/{\text{d}\log R}=-3/2$. It is possible to show that the transport of energy and angular momentum through the propagation of the GI spiral density waves can be divided into two parts \citep{Toomre69,Shu70}: a non-local term and a viscous-like term. Since the non-local term is important only for very high disc-to-star mass ratios, $M_d/M_\star \gtrsim 0.5$ \citep{lodato05}, gravitationally unstable protostellar discs essentially behave as $\alpha$-discs \citep{lodato04, forgan11}.

To characterise the transport of angular momentum through spiral density waves, we need to rely on numerical simulations of gravitationally unstable discs \citep{Cossins09}. Usually in numerical simulations of gravitationally unstable discs, the $\beta$ cooling framework is adopted \citep{Gammie01}. We suppose that the disc is cooling with a rate per unit mass of:
\begin{equation}
    q^{-} = -\frac{e}{t_\text{cool}},
\end{equation}
where $e=c_s^2/\gamma(\gamma-1)$ is the internal energy per unit mass, with the adiabatic index $\gamma = 5/3$, and the cooling time is defined in terms of the dynamical one: $t_\text{cool} = \beta \Omega^{-1}$. In the absence of external heating mechanisms, an initially stable hot disc ($Q>>1$) will cool down, eventually reaching the marginally stable state ($Q=1$). At this point, gravitational instability turns on: the disc develops a spiral structure that, by means of compression and shocks, leads to efficient energy dissipation and heating. In this sense, the Q-stability condition acts as a thermostat so that heating turns on only if the system is sufficiently cold, keeping it in a marginally stable state \citep{KLod16}. In this regime, namely, thermal saturation, the cooling is completely balanced by heating provided by the shocks, and the amplitude of spiral perturbations saturates at a fixed value, according to \cite{Cossins09}:
\begin{equation}\label{beta_sigma}
    \left(\frac{\delta\Sigma}{\Sigma}\right) =\left[\frac{2}{\epsilon \beta \gamma(\gamma-1)} \frac{1}{\mathcal{M}\tilde{\mathcal{M}}} \right]^{1/2}= \chi \beta^{-1/2},
\end{equation}
 {where $\epsilon$ is the heating factor, $\mathcal{M} $ and $\tilde{\mathcal{M}}$ are the Mach numbers relative to the radial phase speed and Doppler-shifted radial phase speed of the wave: respectively, $m\Omega_p/k$ and $m(\Omega -\Omega_p)/k$, with $\Omega_p$ as the spiral pattern frequency. \cite{Cossins09} showed through numerical simulations that the relevant scaling of the spiral density perturbation is $\beta^{-1/2}$ and the other terms of the order of unity. For this reason, we introduce $\chi$, which is of the order of unity.} In the thermal saturation regime, the transport of angular momentum provided by gravitational instability is described within an $\alpha-$framework\citep{KLod16}:
\begin{equation}\label{alpha_beta}
    \alpha_\text{GI} = \left|\frac{\text{d}\log \Omega}{\text{d}\log R}\right|^{-2} \frac{1}{\gamma(\gamma-1)\beta}.
\end{equation}
Combining  Eqs. \eqref{alpha_beta} and \eqref{beta_sigma}, we obtain the relationship between the amplitude of the density wave and the amount of angular momentum 
\begin{equation}
    \frac{\delta \Sigma}{\Sigma} = \frac{3\chi}{2}\sqrt{\gamma(\gamma-1)} \alpha_\text{GI}^{1/2},
\end{equation}
where the last equation is valid for a Keplerian disc. In the following, we assume that the constant of proportionality is $\chi=1$, which is well justified by numerical simulations \citep{Cossins09}.

\subsection{Kinematic signatures of gravitational instability}
Kinematics offers a unique opportunity to quantify the transport of angular momentum in a gravitationally unstable disc. Indeed, when there is a spiral density wave, also the velocity field is affected. \cite{Hall20} predicted that a disc undergoing such instability has clear kinematic signatures in molecular line observations across the entire disc azimuth and radius, called `GI wiggles'. \cite{Longarini21} provided an analytical model to describe such kinematic signatures,  {implemented in the publicly available code \textsc{giggle}\footnote{\url{https://doi.org/10.5281/zenodo.10205110}}}. Under the hypothesis of thin disc, in a marginally unstable regime $(Q=1)$ and in thermal saturation, the amplitude of the velocity perturbation increases with the disc-to-star mass ratio $(M_d/M_\star)$, as has also been described with numerical simulations \citep{Terry22}, and which is proportional to the cooling factor, $\beta^{-1/2}$. However, from the analysis carried out in \cite{Longarini21}, the actual quantity that determines the amplitude of the velocity perturbations $(\delta u_R, \delta u_\phi)$ is the amplitude of the spiral density wave, $\delta\Sigma/\Sigma$. This quantity is intrinsically linked to the efficiency of angular momentum transported by the spiral, which can be described within an $\alpha-$viscosity framework through Eq. \eqref{alpha_beta}. Thus, it is possible to constrain the value of $\alpha_\text{GI}$ from the amplitude of the wiggle. Rewriting the velocity perturbations (Eq. 22 of \cite{Longarini21}) as a function of  $\alpha_\text{GI}$, we obtain:
\begin{equation}
    \delta u_R = 3im\alpha_\text{GI}^{1/2}\sqrt{\gamma(\gamma-1)}\left(\frac{M_d}{M_\star}\right)^2 u_k,
\end{equation}
\begin{equation}
    \delta u_\phi = -\frac{3i\alpha_\text{GI}^{1/2}}{4}\sqrt{\gamma(\gamma-1)}\left(\frac{M_d}{M_\star}\right) u_k. 
\end{equation}
%and thus the amplitude of the wiggle as defined in \cite{Longarini21} scales as
%\begin{equation}
%    \mathcal{A}_w \propto \alpha_\text{GI}^{1/2}.
%\end{equation}

The final expressions for the velocity field are:
\begin{equation}\label{vphi_mod}
    u_\phi(R,\phi) = R\Omega + \Re\left[\delta u_\phi(R)e^{i(m\phi + \psi)}\right]= R\Omega -|\delta u _\phi(R)|\sin(m\phi + \psi),
\end{equation}
\begin{equation}\label{vr_mod}
    u_R(R,\phi) = \Re\left[\delta u_R(R)e^{i(m\phi + \psi)}\right]= -|\delta u _R(R)|\sin(m\phi + \psi).
\end{equation}

\section{Gravitational instability in Elias 2-27}

\subsection{Toomre parameter}

\begin{figure}
    \centering
    \includegraphics[width=\columnwidth]{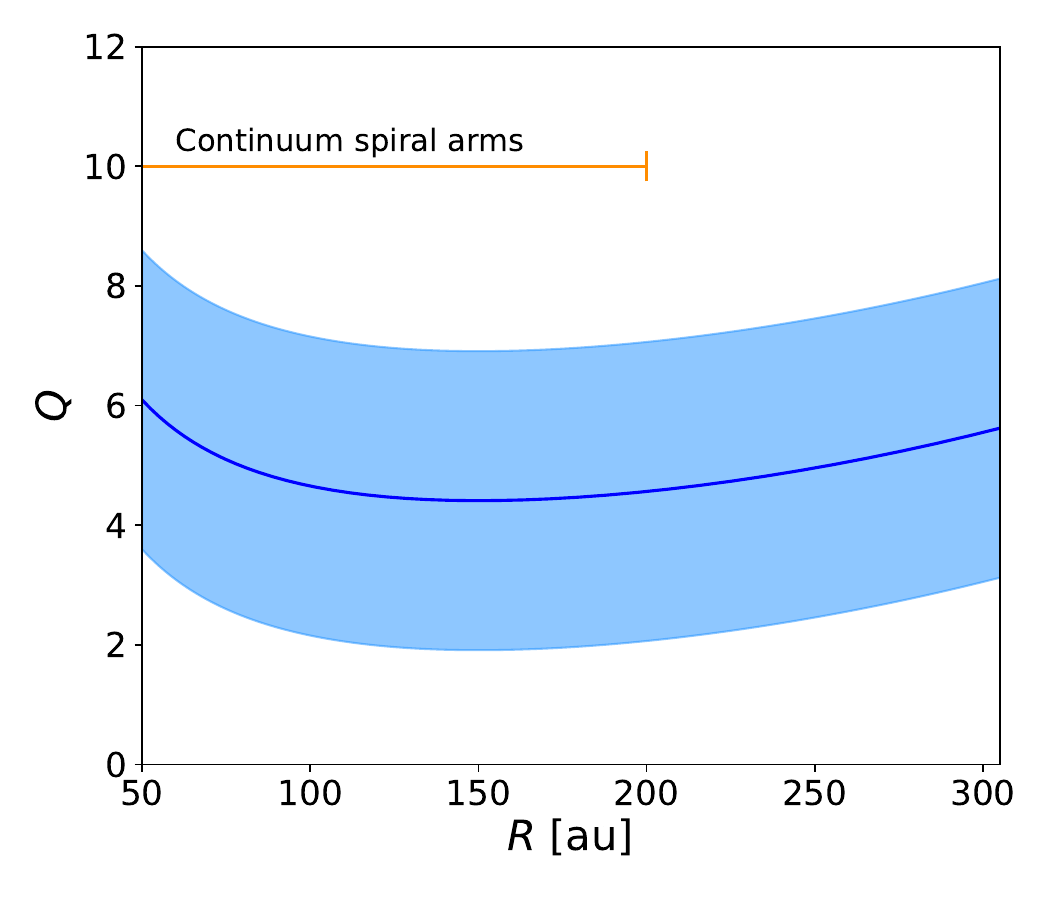}
    \caption{Toomre profile of Elias 2-27, where the shaded region indicates the uncertainties on the disc and star mass from \cite{Veronesi21}.}
    \label{toomre_elias}
\end{figure}

\cite{Veronesi21} estimated the dynamical mass of Elias 2-27 from $^{13}$CO and C$^{18}$O rotation curves, and found $M_\star = 0.46\text{M}_\odot\pm 0.03$ and $M_d=0.08\pm0.04\text{M}_\odot$.  {In their fit, they used a self-similar surface density given by}
\begin{equation}
    \Sigma = \frac{M_d}{2\pi (200\text{au})^2}\left(\frac{R}{200\text{au}}\right)^{-1} \exp\left[-\frac{R}{200\text{au}}\right],
\end{equation}
 {where the scale radius has been fixed to $R_c=200\text{au}$. As for the thermal structure, they assumed a vertically isothermal disc with $T(R) = 20 (R/60\text{au})^{0.5}$K taken from \cite{Perez16}, which corresponds to a sound speed of $c_s = 281\text{m/s}(R/60\text{au})^{-0.25}$.} With this information, it is possible to compute the Toomre parameter profile for Elias 2-27, as displayed in Fig. \ref{toomre_elias}. Despite not being exactly $Q=1$, the Toomre profile is close enough to the critical threshold to consider gravitational instability to be significant. In addition, we are not considering uncertainties on $T(R)$, that can impact on the $Q-$parameter estimate.  {Finally, we know that the disc shows non-axisymmetric features in gas and dust, making the azimuthally averaged Toomre profile solely an approximation of the actual value. }

\subsection{Dataset}
In this work, we use the $^{13}$CO $J=3-2$ datacube presented in \cite{paneque21}. The images have been obtained with a robust parameter of 0.5, resulting in a beam size of $0.26^{\prime\prime}\times0.25^{\prime\prime}$ for the $^{13}$CO and a spectral resolution of $\Delta v =111$m/s. Further details of the observations and reduction can be found in \cite{paneque21}. In this work, we use a Gaussian velocity map obtained with \textsc{bettermoments} \citep{bettermoments}. The code also returns a map of the errors on the velocity field \citep{teague19}. In the analysis, we masked the emission coming from the first two beams ($\sim50\text{au}$).

\subsection{Model of the GI wiggle of Elias 2-27}\label{section_wiggle_elias}
In the analytical model for the GI wiggle of \cite{Longarini21}, the amplitude of the velocity perturbations is determined by the disc to star mass ratio and the cooling factor. There is a degeneracy between the two quantities; however, in the case of Elias 2-27 the value of the disc to star mass ratio is known \citep{Veronesi21}. \cite{Longarini21} showed that the amplitude of the wiggle in the PP (position-position) space scales as $\beta^{-1/2}$ (i.e. $\alpha_\text{GI}^{1/2}$), and they proposed this relationship as a way to constrain this unknown parameter. In this paragraph, we study the wiggle in the PV (position-velocity) space, as done in \cite{speedie24}.
%, and we constrain the cooling. In appendix \ref{app_1}, we compare the results in the PP space.
We consider the observed velocity field: 
\begin{equation}
    u_\text{obs}(R,\phi) = \left[u_\phi(R,\phi)\cos\phi + u_R(R,\phi)\sin\phi\right]\sin i,
\end{equation}
where $u_\phi$ and $u_R$ are described by the model presented in Eqs. \eqref{vphi_mod} and \eqref{vr_mod} and $i$ is the disc inclination. We compute the variation of velocity along the semi-minor axis of the disc (i.e. $\phi = -\pi/2$), because along this axis only the radial velocity contributes to the observed velocity field
\begin{equation}
    u_\text{obs}(R,-\pi/2) = -2m\beta^{-1/2}\left(\frac{M_d(R)}{M_\star}\right)^2u_k\sin\left(\psi(R) + \psi_0\right),
\end{equation}
where $\psi(R)$ is the phase function of the spiral that is given by
\begin{equation}
    \frac{\text{d}\psi}{\text{d}R} = \frac{m}{R\tan\alpha_p},
\end{equation}
with $\alpha_p$ being the pitch angle of the spiral and $\psi_0$ is just a phase shift. We suppose that the pitch angle is constant over the radial extent of the disc, that is well justified for GI spirals \citep{Cossins09}. We note that the last equation can be written as a function of $\alpha_\text{GI}$ as:
\begin{equation}\label{uobs_model}
    u_\text{obs}(R,-\pi/2) = -3m\alpha_\text{GI}^{1/2}\sqrt{\gamma(\gamma-1)}\left(\frac{M_d(R)}{M_\star}\right)^2u_k\sin\left(\psi(R) + \psi_0\right).
\end{equation}

\begin{figure}
    \centering
    \includegraphics[width=\columnwidth]{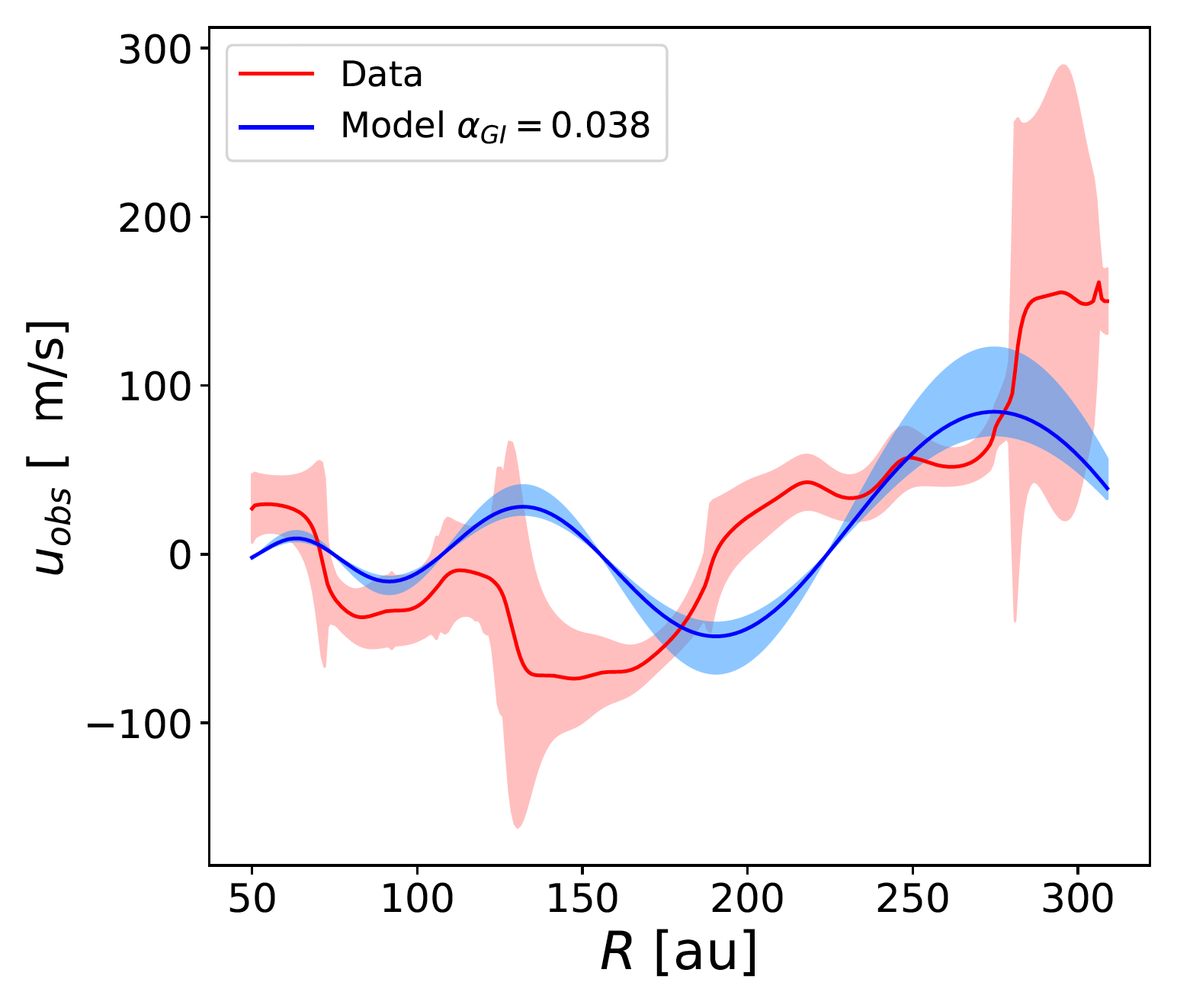}
    \caption{GI wiggle in the PV space: comparison between the the data (red line) and the model (blue line). The shaded region represents the error on the model, driven by the the uncertainties on the star and disc masses from \cite{Veronesi21}.}
    \label{PV_wiggle}
\end{figure}

We then extract the PV wiggle by cutting along the semi minor axis and considering just the southern part of the disc, on account of cloud contamination \citep{paneque21}. The errors on the velocity are returned by  \citep{bettermoments} as shown in figure \ref{PV_wiggle}. The disc and spiral parameters we use are taken from literature, namely, $M_\star = 0.46\text{M}_\odot$, $M_d = 0.08\text{M}_\odot$, $R_c=200$au \citep{Veronesi21}, $\alpha_p = 13^\circ$, $i= 56.2^\circ$, $\text{PA}=118.8^\circ$ \citep{paneque21}, and $m=2$. Hence, the only free parameters in Eq. \eqref{uobs_model} are $\alpha_\text{GI}$ and $\psi_0$. To fit the curve to the data, we used the method of nonlinear least squares implemented in \textsc{Scipy} \citep{scipy2020}.  {While the spiral pattern in the continuum extends to approximately $200\text{au}$, non-axisymmetric kinematic signatures in the gas emission are observable across the entire radial extent of the disk. Hence, in the fitting procedure, we analysed the signal up to the outer edge of the disc.} The best-fit values are $\alpha_\text{GI}=0.038\pm0.018$ and $\psi_0 = 43^\circ\pm 1 ^\circ$. The value of $\alpha_\text{GI}$ corresponds to a $\beta = 10.5$, meaning that the cooling time of the system is approximately ten times the dynamical one. We estimated the error on $\alpha_\text{GI}$ propagating the uncertainties on star and disc masses\footnote{The fitting procedure implemented in \textsc{Scipy} returns an error for the best fit parameters. As for the $\alpha_\text{GI}$ parameter, the error provided by the fit is subdominant compared to the one driven by the star and disc masses uncertainties.}. 

Figure \ref{PV_wiggle} shows the comparison between the extracted PV wiggle from $^{13}$CO data and the model that best describes the data. By comparing the model and the data, we observe that the overall shape of the perturbation is well reproduced. However, it appears that there is a radial shift between the two curves.  {This effect can be attributed to the fact that in the analytical model, we assume that a single spiral mode is present, with $m=2$, and we also suppose that the perturbation wave-number, $k,$ is the most unstable $(k_\text{uns})$. While the $m=2$ mode, with $k=k_\text{uns}$ may be the dominant one, determining the overall morphology of the spiral, this does not prevent the presence of additional lower amplitude modes that will interfere with the dominant one and thus create a more complex pattern than the purely sinusoidal one that we assume here. The scope of this work is to find the amount of stress generated by gravitational instability; hence, we are interested in the amplitude of the wiggle, not in the whole shape. For this reason, the additional lower amplitude modes are not a concern in terms of the amplitude of the perturbation.}

\begin{figure*}
    \centering
    \includegraphics[width = \textwidth]{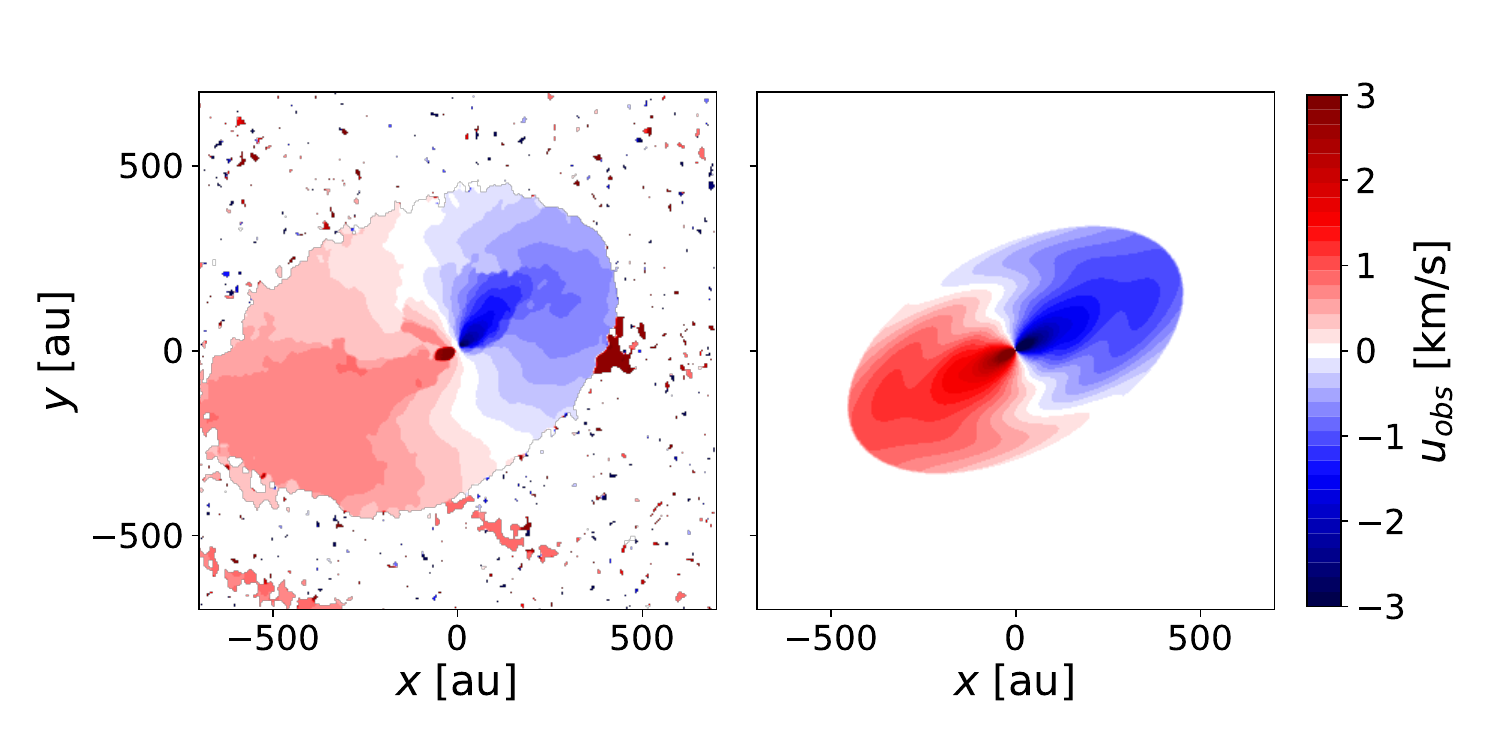}
    \caption{Comparison between Elias 2-27 velocity map (left) and the best fit model for the velocity field with $\alpha_\text{GI}=0.038$ (right).}
    \label{M1_analytical}
\end{figure*}

 {Figure \ref{M1_analytical} shows the analytical observed velocity field with the aforementioned parameters. Despite its simplicity, the analytical model matches the shape of the different channels of the data. In particular, the bending of the isovelocity contours is described very well by the analytical model, especially in the blue-shifted region, where the effect of cloud contamination are negligible.}
%This effect can be imputed to the hypothesis of constant pitch angle of the spiral. As a matter of fact, a radial variation of the opening angle could create a shift on the position of the maxima-minima of the perturbation. 
The analytical models are produced using the publicly available code \textsc{giggle}. It is important to point out that the model presented in \cite{Longarini21} is limited to two dimensions and does not consider the vertical extent of the disc. There is evidence suggesting that the $^{13}$CO emission in Elias 2-27 is optically thick \citep{paneque22}; consequently, the received signal does not originate from the midplane, but from a layer at $z\neq 0$. The analytical model we employ is not able to reproduce this effect, and, for a proper comparison, hydrodynamical and radiative transfer simulations would be needed. In any case, assuming that $z/R<0.3$ for the $^{13}$CO \citep{paneque22}, the difference in inclination induced by the finite height of the emitting layer is $\delta i = \arcsin(0.3)\simeq 0.29$. The corresponding geometrical error is roughly $1-\sin(i+\delta i)/\sin(i)\simeq 15\%$, that is subdominant compared to the one driven by the mass estimate.

%While the model cannot reproduce this aspect, it is expected that this would have minimal impact on the wiggle estimate, given that $z/R<0.4$ across the entire radial extent of the disc \citep{paneque22}. 

\subsection{Angular momentum transport and accretion}
For a gravitationally unstable disc, kinematics offers a unique opportunity to quantify the transport of angular momentum of a gravitationally unstable disc. We show that gravitational instability in Elias 2-27 transports angular momentum and the equivalent $\alpha-$viscosity parameter is $\alpha_\text{GI} = 0.038\pm 0.018$. Since the effective viscosity is responsible for the accretion process, it is useful to predict the expected accretion rate onto the central object, and compare it to the observed one. 

According to the self-similar solution \citep{selfsimilar}, the surface density and the accretion rate of the disc can be written as:
\begin{equation}
    \Sigma=\frac{M_d}{2\pi R_c^2}\left(\frac{R}{R_c}\right)^{-1} \exp\left[-\frac{R}{R_c}\right],
\end{equation}
\begin{equation}\label{mdot_SS}
    \dot{M} =  \frac{3M_d\nu_c}{2R_c^2}\exp\left[-\frac{R}{R_c}\right] \left(1-\frac{2R}{R_c}\right),
\end{equation}
where we have supposed that $\nu = \nu_c (R/R_c)$, where the subscript $c$ means that the quantity is evaluated at the scale radius $R_c$. Within a $\alpha-$viscosity framework, the kinematic viscosity, $\nu,$ is
\begin{equation}
    \nu = \alpha c_s H = \alpha \left(\frac{H}{R}\right)^2 u_k R.
\end{equation}
We note that within our assumption that $T\propto R^{0.5}$, for a constant $\alpha$ the kinematic viscosity scales as $\nu \propto R$.
In this way, we can write the accretion rate onto the central object as the limit for $R\to 0$ of Eq. \eqref{mdot_SS}:
\begin{equation}\label{mdot_star}
    \dot{M}_\star =- \frac{3\alpha}{2}\left(\frac{H}{R}\right)_{R_c}^2M_d\Omega_c,
\end{equation}
where $\Omega_c = \Omega(R_c)=\sqrt{GM_\star/R_c^3}$.

Using $\alpha_\text{GI}=0.038$ and the disc parameters described in the previous paragraph, it is possible to compute the accretion rate onto the central object by using Eq. (\ref{mdot_star}). Thus, we obtain:
\begin{equation}
    \log_{10}\dot{M}_\star [\text{M}_\odot/\text{yr}] =-6.99\pm 0.17,
\end{equation}
where the error has been computed through propagation from the errors in $\alpha_\text{GI}$. The model for the accretion rate of Elias 2-27 reproduces the one measured by \cite{natta06} very well, namely,  $\log_{10}\dot{M}_\star [\text{M}_\odot/\text{yr}] =-7.2\pm 0.5$. They used J and K-band spectra to derive the mass accretion rate of objects in the $\rho-$Ophiuchi star forming region from the intensity of the hydrogen recombination lines. More recently, a new estimate of Elias 2-27 accretion rate was provided by \cite{testi22}, being $\log_{10}\dot{M}_\star [\text{M}_\odot/\text{yr}] =-7.3$, which is consistent with the previous measurement.

 {To constrain $\alpha_\text{GI}$ in Elias 2-27, we made the strong hypothesis that this quantity, or equivalently the cooling $\beta$, is constant throughout the disc. We are aware that, realistically, this is not the case. However, we point out that our disc model is self consistent. Indeed, assuming a self similar solution for the surface density with $\Sigma \propto R^{-1}$, we are imposing that the kinematic viscosity scales as $\nu \propto R$. Since $\nu = \alpha c_s H$, with $c_s\propto R^{-0.25}$ \citep{Perez16} and $H\propto R^{1.25}$, the viscosity coefficient, $\alpha,$ should be constant with radius. We also point out that, in order to obtain an estimate of the amount of angular momentum transported within the disc from the GI wiggle, an assumption should be made with respect to how $\alpha$ varies with the radius. A more realistic disc model for Elias 2-27 is not within the scope of this work. We stress that the choice of a self-similar profile for the surface density was made to maintain consistency with the work of \citet{Veronesi21}.}

\section{Discussion}
The ability of our model to correctly reproduce the observed accretion rate points to the fact that gravitational instability is responsible for angular momentum transport in this system. As a matter of fact, the value of the $\alpha-$viscosity we get from the GI wiggle is the one required to explain the observed accretion rate onto the central object, having fixed the density structure of Elias 2-27. In addition, the strong hypothesis we made is that viscous processes are responsible for accretion.  The inferred value for the $\alpha-$viscosity is higher than usually assumed $(\sim10^{-3}-10^{-4})$. This is not surprising, since the strength of the viscosity generated by gravitoturbulent motions is higher than the expected in the non-self gravitating state \citep{Cossins09}.

\subsection{Infall and interaction with the environment}
Elias 2-27 is a young system, and its interactions with the surrounding environment are possibly perturbing the disc. Specifically, the disc is partially embedded within the molecular cloud, which absorbs $^{12}$CO and a portion of the $^{13}$CO emission \citep{Perez16, paneque21}, then feeds the disc with mass. In particular, infall is an alternative way to trigger gravitational instability \citep{kratter06,kratter08,KLod16}. Gravitational instability occurs when the Toomre parameter is of the order of unity, and this threshold can be reached by cooling the disc (i.e. decreasing the sound speed) or by adding mass (i.e. increasing the surface density). When GI is triggered by infall, there is a mechanism akin to the thermal saturation \citep{KLod16}. 

Our $\alpha_\text{GI}$ estimate remains agnostic to the source of self-regulation, whether it arises from cooling or from the addition of mass to the disc. Indeed, what we are measuring through the wiggle is the amplitude of the surface density perturbation, $\delta\Sigma/\Sigma$. It is intrinsically linked to the efficiency of angular momentum that is transported by the spiral, regardless of the origin of the instability.

\subsection{Planet formation in Elias 2-27}
The value of $\beta$ we measure is much higher than the threshold for disc fragmentation into bound gas clumps \citep{Gammie01, deng17}, implying that (while the disc is gravitationally unstable) direct planet formation through gravitational instability is unlikely.  {However, many studies \citep{paardekooper12,young15} showed that stochastic fragmentation of gas in spiral arms can happen for high values of $\beta$.} As for the solid component, \cite{longarini23b,longarini23a,rowther24}  investigated the possibility of forming planetary cores in gravitationally unstable discs through dust collapse. They found that for a sufficiently long cooling time, $\beta>10,$ and high disc-to-star mass ratio, $M_d/M_\star \sim 0.2$, dust efficiently collects inside spiral arms and its dispersion velocity is so low to induce collapse into bound objects with a mass of $\sim 10\text{M}_\oplus$. The inferred disc-to-star mass ratio \citep{Veronesi21} and the cooling time for Elias 2-27 make it a perfect candidate for planet formation through dust collapse. 

\cite{Dsharp3} characterised annular substructures in the discs within the DSHARP sample and found that Elias 2-27 has a gap at $R_g = 69.1\pm 0.4$au with a width of $\Delta = 14.3\pm 1.1$au. Although several mechanisms can explain the origin of gaps in protoplanetary discs, a common explanation is planet disc interaction. Under the planetary interpretation, the width of the gap scales as the Hill radius of the planet, defined as:
\begin{equation}
    R_h = \left(\frac{M_p}{3M_\star}\right)^{1/3}R_g,
\end{equation}
where $M_p$ is the mass of the protoplanet. The last relation has been obtained by averaging results from hydrodynamical simulations. Following \cite{lodato19}, the relation between the gap width and the Hill radius is $\Delta = 5.5R_h$, that translates into 
\begin{equation}
    M_p = 3\left(\frac{\Delta}{5.5R_g}\right)^3 M_\star.
\end{equation}
Using the gap width and location of \cite{Dsharp3}, and the star mass of \cite{Veronesi21}, the inferred mass of the protoplanet is $M_p = 24\pm 6\text{M}_\oplus$. This result is in good agreement with the mass range of \cite{longarini23b,longarini23a}. Another element that points towards the dust collapse is the value of the Toomre parameter. As shown in \cite{longarini23b}, when the gravitational instability is driven by the cold component (dust in this case), the critical value of the Toomre parameter is $>1$, as observed in Elias 2-27 (see Fig. \ref{toomre_elias}).

\section{Conclusion}
In this work, we investigate the kinematic signatures of gravitational instability in the protoplanetary disc Elias 2-27. It is well known that gravitational instability leaves clear kinematic perturbations in molecular line emission \citep{Hall20} and their characteristics are related to the spiral density wave \citep{Longarini21}. There are multiple arguments suggesting that Elias 2-27 is undergoing gravitational instability. Under the hypothesis that angular momentum is transported through the GI spirals, we estimate the $\alpha-$viscosity of the system, and link it to the accretion rate onto the central object. We find $\alpha_\text{GI} = 0.038\pm0.018$ and $\log_{10}\dot{M}_\star = - 6.99\pm0.17\text{M}_\odot / \text{yr}$. There is a very good agreement between the observed accretion rate and the one estimated from our model, pointing to the fact that gravitational instability is at play in this system and that is indeed driving angular momentum transport.  {We underline that the results obtained in this work are valid assuming a disc model (as described in Sect. \ref{section_wiggle_elias}) and by fitting for the amplitude of the velocity perturbation in the central channel of the velocity map.} The range of disc masses \citep{Veronesi21} and cooling factors inferred by our model makes Elias 2-27 a perfect candidate for dust collapse and the formation of planetary cores in spiral arms. The gap present in dust continuum emission at $\sim70$au points to the presence of a $\sim20\text{M}_\oplus$ protoplanet, in agreement with the mass range of planets formed by collapse of the dust component by \cite{longarini23a}.

\begin{acknowledgements}
This work has received funding from the European Union’s Horizon 2020 research and innovation programme under the Marie Sklodowska-Curie grant agreement \# 823823 (RISE DUSTBUSTERS project). CL and CJC have been supported by the UK Science and Technology research Council (STFC) via the consolidated grant ST/W000997/1. PC acknowledges support by the Italian Ministero dell Istruzione, Universit\`a e Ricerca through the grant Progetti Premiali 2012 – iALMA (CUP C52I13000140001). J.S. acknowledges financial support from the Natural Sciences and Engineering Research Council of Canada (NSERC) through the Canada Graduate Scholarships Doctoral (CGS D) program. The authors thank Francesco Zagaria, Andrew Sellek and Myriam Benisty for useful discussions.
\end{acknowledgements}

\bibliographystyle{aa} % style aa.bst
\bibliography{bibliography.bib} % your references Yourfile.bib
\nocite{*}

%\appendix
%\section{Wiggle in the PP space}\label{app_1}
%In this appendix we compare the results of our analysis in the PP space, as previously outlined in \citep{Longarini21}. We extract the isoveocity contour $u_\text{obs} = u_\text{sys}$ and we consider only the contribution coming from the southern part of the disc, as explained in section \ref{section_wiggle_elias}. We compute the amplitude of the wiggle as presented in \cite{Longarini21}, and we verify that the value of $\alpha_\text{GI}=0.038$ correctly reproduces the amplitude in the PP space of $\mathcal{A} = ...$.

%\begin{figure}
%    \centering
%    \includegraphics[width = %\columnwidth]%{images/wiggle_PP_mod.pdf}
%    \caption{GI wiggle in the PP space: comparison between the $^{13}$CO data (red line) and the model with $\beta=10.5$ (blue line).}
 %   \label{wiggle_PP}
%\end{figure}

\end{document}